\documentclass[prl,twocolumn,amsmath,amssymb,citeautoscript,longbibliography]{revtex4-2}
\usepackage[unicode=true,bookmarks=true,bookmarksnumbered=false,bookmarksopen=false,breaklinks=false,pdfborder={0 0 0},pdfborderstyle={},backref=false,colorlinks=true,]{hyperref}
\hypersetup{linkcolor=blue,citecolor=blue,urlcolor=blue}
\usepackage{lipsum}
\usepackage{amsfonts}
\usepackage{mathtools}
\usepackage{graphicx} 
\usepackage[]{color}

\date{May 2025}

\begin{document}

\title{Spontaneous emission as a bridge from Lindbladian to nonreciprocal reservoirs} 

\author{C. J. Bolech}
\altaffiliation{\textit{Department of Physics, University of Cincinnati, 345 Clifton Court, Cincinnati, Ohio 45221-0011, USA}}
\affiliation{University of Geneva, DQMP, 24 quai Ernest-Ansermet, 1211 Geneva, Switzerland}

\author{T. Giamarchi}
\affiliation{University of Geneva, DQMP, 24 quai Ernest-Ansermet, 1211 Geneva, Switzerland}

\begin{abstract}
We study an out-of-equilibrium quantum system in which a state connecting two reservoirs is also coupled by stimulated and spontaneous emission of photons to an antitrapped state, thus implementing particle loss. After revisiting the spontaneous emission process, we show that the proper effective description of such a system requires one to go beyond the usual Lindbladian formalism and includes a nonreciprocal (``non-Hermitian'') coupling to the reservoir modeling the untrapped state. The presence of both, the reservoirs and the nonreciprocal coupling, have observable consequences that we compute, for example, by looking at the quantum Zeno effect in the loss current. We discuss the connection of our findings to possible experiments in cold atomic gases.
\end{abstract}

\maketitle


Although the hermiticity of operators associated with physical observables is one of the basic postulates of quantum theory, the use of effective non-Hermitian formalisms to describe open quantum systems, resonances, and hybrid quantum-classical systems has already a well established history \cite{Bender2007,Moiseyev_NonHermitianQM,ElGanainy2018,AshidaGongUeda2020} that interrelates to other approaches like Lindbladians or quantum master equations \cite{CohenTanoudji_ProcessusPhotonsAtomes,BreuerPetruccione_OpenQuantumSystems,Kamenev2023}.
A well known example is the nonreciprocal (or \emph{asymmetric}) hopping model of Hatano and Nelson \cite{HatanoNelson1996,*HatanoNelson1997}, which brings into a quantum context the classical non-equilibrium statistical physics of \emph{asymmetric simple exclusion processes} (ASEP) \cite{Spitzer1970,Derrida1992} and has received a lot of attention recently in connection to realizations using systems with tunable gain and loss \cite{McDonald2022,Liu2022,Gao2023,Oersel2025,Tao2025}.

Meanwhile, the study of emergent collective behavior was bolstered by advances in the realization of macroscopic coherent states such as Bose-Einstein condensates \cite{Pethick2008,pitaevskii_becbook} and the subsequent development of cold atomic systems in optical lattices both bosonic and fermionic \cite{bloch_cold_lattice_review,esslinger_annrev_2010}, systems in cavities \cite{ritsch_cavity_review}, and photonic and mixed systems \cite{Carusotto2013}. Moreover, these systems have allowed the realization of out-of-equilibrium situations driven by loss of particles via collision with an electron beam \cite{BarontiniOtt_LocalizedDissipation}, decay via untrapped states \cite{Corman_DissipativeAtomicPointContact,zhao_losses_TLL} or pumping and losses \cite{kasprzak_BEC_polaritons}. These experimental
realizations have naturally triggered a flurry of corresponding theoretical activity.

An important natural question is the precise formalization of the ``non-Hermitian'' description needed for a given experimental realization. Most open systems can be modeled by a Lindbladian description or the contact with a reservoir which, when traced, \cite{Gardiner_QuantumNoise,sieberer_keldysh_lindblad_review} gives the corresponding irreversibility from purely Hermitian processes. The use of non-Hermitian ``Hamiltonians'' is usually justified in the cases of post-selection of trajectories which do not undergo quantum jumps \cite{ashida_review_nonhermitian,BoschAguilera2022}, outgoing boundary conditions \cite{Moiseyev_NonHermitianQM}, or explicit breaking of symmetry for other reasons \cite{KakashviliBolech_TimeLoop,Pan_NonHermLinearResp}.
Another point is when these two descriptions could be related (for example, the asymmetric couplings in the quantum version of ASEP being modeled by the presence of reservoirs \cite{McDonald2022}).

In this Letter, we show by means of an example a route by which \emph{asymmetric terms} can arise directly in a quantum-mechanical context as a description of the full dynamics of an open system (involving neither transients nor post-selection). 
We focus on the electronic transitions between atomic levels of a single atom (ion or molecule) mediated by the interaction with light modes. We distinguish between \emph{stimulated} and \emph{spontaneous} transitions and revisit the way to model the latter. 
An effective \emph{nonreciprocal} coupling to a Markovian bath arises naturally from the full open-system description.
In addition, we argue that the degree of asymmetry is tunable and has experimental signatures discoverable in non-equilibrium transport setups.


\begin{figure}[t]
\begin{center}
\includegraphics[width=\columnwidth]{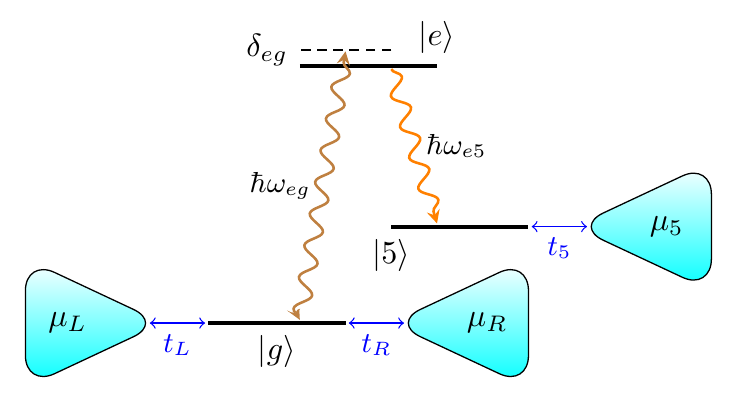}
\end{center}
\caption{\label{Fig:energy-level-diagram}
Schematic depiction of the experimental setup. Transport between two atomic Fermi-gas reservoirs with chemical potentials $\mu_\textsc{l}$ and $\mu_\textsc{r}$ goes through a narrow constriction and is diverted by a narrowly focused laser beam that triggers a lambda-system transition. Atoms that relax into the secondary ground state are not trapped and quickly leave the system, modeled as a \emph{loss current} into a third reservoir with chemical potential $\mu_5$. Such a composite system of fermions and photons can be mapped onto an effective purely fermionic nonreciprocal (``non-Hermitian'') model as shown in Fig.~\ref{Fig:effective-model}.}
\end{figure}

We consider a configuration inspired by recent experiments \cite{Corman_DissipativeAtomicPointContact,HuangMohanVisuri_DissipativeQPC}, in which a system is stimulated into a state that decays by spontaneous emission to an antitrapped state (see Fig.~\ref{Fig:energy-level-diagram}). 
This situation, referred to as a Lambda system, allows to engineer loss of particles in a state which serves to connect two fermionic reservoirs to perform transport experiments. 
We shall use that particular example to construct a generic minimal model that captures the differences between spontaneous and stimulated transitions and allows to generate actions with nonreciprocal terms, without the need of post-selection of the data. 
The Hamiltonian consists of an atomic part, a photon part, and some light-matter interaction terms;
\begin{equation}
H=H_{\mathrm{at}}+H_{\mathrm{ph}}+H_{\mathrm{int}}    
\end{equation}
We consider a degenerate Fermi gas of $^{6}$\textrm{Li} atoms in a harmonic trap. Unlike in the quoted experiments, we do not include a {\textquotedblleft spin\textquotedblright} degree of freedom and take all atoms to be in the (first) hyperfine ground state, $\left\vert g\right\rangle $. 
The dissipation scheme has the relevant atomic energy state $\varepsilon_{g}$ optically excited by the \emph{dissipation} probe laser to an excited state $\left\vert e\right\rangle $ of energy $\varepsilon_{e}$ which decays predominantly (over a 99\% probability) to an auxiliary ground state $\left\vert 5\right\rangle$, (the fifth hyperfine ground state) of energy $\varepsilon_{5}>\varepsilon_{g}$, which quickly leaves the system due to photon recoil \cite{HuangMohanVisuri_DissipativeQPC}; see Fig.~\ref{Fig:energy-level-diagram}.
In second-quantized language, we write
\begin{equation}
H_{\mathrm{at}}=\varepsilon_{g}\,c_{g}^{\dagger}c_{g}+\varepsilon_{e}%
\,c_{e}^{\dagger}c_{e}+\varepsilon_{5}\,c_{5}^{\dagger}c_{5}%
\end{equation}
There are no direct transitions between these two ground states, but they are connected via $\left\vert e\right\rangle$ with a two-photon process.
After shifting away their vacuum energy, we write for the photons,
\begin{equation}
H_{\mathrm{ph}}=\hbar\omega_{eg}\,a_{eg}^{\dagger}a_{eg}+\hbar\omega
_{e5}\,a_{e5}^{\dagger}a_{e5}    
\end{equation}
We work in the Coulomb gauge and assume that the static Coulomb potential was included in the resolution of the atomic energy levels.
The remaining transverse modes of the electromagnetic field couple to the atoms via the minimal substitution.
Because the atoms are non-relativistic (have a quadratic dispersion relation), there are in principle both one- and two-photon terms in the matter-light coupling description. 
Only the former are relevant for the $g\leftrightarrow e$ transition because it is driven by a laser of the appropriate frequency. 
On the other hand, the $e\rightarrow5$ transition is not driven and spontaneous emission could happen in both channels, but, for simplicity and following the experimental phenomenology, we will model it as well as a one-photon process
\footnote{We will not concern ourselves with momentum conservation (that can be absorbed easily by the center-of-mass motion of the atoms and produces their recoil that we include separately) and will focus entirely on energy considerations.}. 
We thus write:
\begin{equation}
H_{\mathrm{int}}=\lambda_{eg}\,c_{e}^{\dagger}a_{eg}c_{g}+\lambda_{e5}%
\,c_{e}^{\dagger}a_{e5}c_{5}+\mathrm{h.c.}    
\end{equation}
with $\lambda_{e5}\gg\lambda_{eg}$ according to the parameters in our guiding experimental example
\footnote{This interaction can also be arrived at by chaining together two Jaynes-Cummings models \cite{Jaynes1963} that share the middle state $\left\vert e\right\rangle$ and carrying out the usual rotating-wave approximation to neglect the quickly oscillating \textit{counter-rotating} terms.}.


For the stimulated transition, $g\leftrightarrow e$, the description we have presented so far should be adequate. 
The only additional consideration to add is that due to the laser beam the population of the $\omega_{eg}$ photon mode should be taken to be large and the difference between adding or removing one photon can be safely neglected. 
This is consistent with saying that the electromagnetic mode is in a photonic coherent state.

For the spontaneous transition, $e\rightarrow5$, the situation is more nuanced. 
In simple descriptions, one might consider a Hilbert space with only zero or one photons (or a similar small number), as is sometimes done in the modeling of radioactive decay.
In the engineered-dissipation setting considered here, we want to describe a continuous process rather than a series of statistical one-off events. 
A plausible way to do that is by considering a standing photon mode (as in the stimulated case) but in a leaky cavity, with a leakage that can be taken to be as large as needed to match the desired decay rate; we shall explore this route systematically, using a \emph{Lindblad master equation} approach \cite{GoriniKossakowskiSudarshan_Linblad,Lindblad1976}
\footnote{An alternative could be to impose \emph{outgoing boundary conditions} \cite{Moiseyev_NonHermitianQM} on the generated $\hbar\omega_{e5}$ photons, which could be argued would yield directly a non-Hermitian asymmetric atom process.}.


Let $\rho$ be the reduced density matrix of a \emph{system} (obtained after tracing out its \emph{environment}).
If the environment is large, evolves relatively fast and is weakly coupled, one can make a Born-Markov approximation
\footnote{sometimes a rotating-wave approximation for the system is also invoked, but it can be replaced by a modified Markov approximation in a system-independent way \cite{NathanRudner_UniversalLindblad}.} 
and say that the \emph{open} system evolves according to a Lindblad quantum master equation (in its diagonal form) \cite{BreuerPetruccione_OpenQuantumSystems},
\begin{equation*}
\mathrm{i}\hbar\partial_{t}\rho=\left[  H,\rho\right]  +\mathrm{i}\hbar
\sum_{m}\gamma_{m}\left(  L_{m}\rho L_{m}^{\dagger}-\frac{1}{2}\left\{
L_{m}^{\dagger}L_{m},\rho\right\}  \right)    
\end{equation*}
where $\left[{_\square},\rho\right]$ and $\left\{{_\square},\rho\right\}$ are the commutator and anticommutator, respectively. 
This is the most general local linear evolution rule that respects the basic physical properties of the reduced density matrix (hermiticity, positive definiteness, and normalization).
The $L_{m}$ are known as Lindblad jumps or leap operators and describe the nature of the different dissipative channels 
\footnote{It is well known that considering a Lindblad evolution with single-particle \textit{L}-operators is equivalent to considering single-particle tunneling \emph{into empty} (or \emph{from full}) Landauer-B{\"{u}}ttiker leads and taking the chemical potential to minus (or plus) infinity in order to drop the temperature and frequency dependence from the thermal distribution functions of those baths; cf.~Ref.~\onlinecite{Jin_MarkovianReservoirs}.}\!
\footnote{For sufficiently short times, one might be able to neglect the $L_{m}\rho L_{m}^{\dagger}$ terms (quantum-jump contributions) to match experiments after \emph{post-selection} of the data. In that case, the von Neumann equation of motion for $\rho$ turns to $\mathrm{i}\hbar\partial_{t}\rho=(H_{\mathrm{eff}}\rho-\rho H_{\mathrm{eff}}^{\dagger})$ with $H_{\mathrm{eff}}=H-\mathrm{i}\frac{\hbar}{2}\sum_{m}\gamma_{m}L_{m}^{\dagger}L_{m}$. 
This is practical for times shorter than the smallest $1/\gamma_{m}$ \cite{YamamotoUedaKawakami_NonHermitian} and can be extended to longer times using quantum-trajectory techniques that do Monte Carlo sampling over quantum jumps.
In our study, we are instead interested in times that are long enough that a steady state is reached.}.

To proceed further, it is useful to consider the Lindblad evolution of expectation values \cite{BarontiniOtt_LocalizedDissipation} or multi-point correlators \cite{Witthaut,AlbaCarollo} given by the standard trace formula $\left\langle O\right\rangle =\operatorname{Tr}\left(  O\rho\right)$.
Working in the Schr\"{o}dinger picture, we first consider time-independent operators $O_{L}$ such that $\left[  O_{L},L_{m}^{\left[  \dagger\right]  }\right]  =0$ $\forall m$ and find that
\begin{equation}
\mathrm{i}\hbar\partial_{t}\left\langle O_{L}\right\rangle =\operatorname{Tr}%
\left(  O_{L}\mathrm{i}\hbar\partial_{t}\rho\right)  =\operatorname{Tr}\left(  O_{L}\left[  H,\rho\right]  \right)    
\end{equation}
In other words, the dynamics of these expectation values is determined solely by $H$ and the \textit{L}-operators do not contribute. (In particular, for $O_{L}=\mathbb{I}$ one finds that $\operatorname{Tr}\left(\partial_{t}\rho\right)=0$.)

Next, we move on to the case of the expectation value of an \textit{L}-operator. We assume that it commutes with the \textit{L}-operators in all the other dissipation channels and drop the index $m$, (but in general $\left[  L,L^{\dagger}\right]$ can still be nonzero). 
After a little algebra, we have
\begin{equation}
\mathrm{i}\hbar\partial_{t}\left\langle L\right\rangle
=\operatorname{Tr}\left(  L\left[
H_{\mathrm{eff}},\rho\right]  \right)
\end{equation}
where $H_{\mathrm{eff}}\equiv H-\mathrm{i}\frac{\hbar\gamma}{2}L^{\dagger}L$, while
\begin{equation}\label{Eq:Ldagger}
\mathrm{i}\hbar\partial_{t}\left\langle L^{\dagger}\right\rangle
=\operatorname{Tr}\left(  L^{\dagger}\left[  H_{\mathrm{eff}}^{\dagger}%
,\rho\right]\right)
\end{equation}
Since $\rho$ is Hermitian, these two equations of motion are related by Hermitian conjugation.

Back to the $e\rightarrow5$ spontaneous-decay transition, we expect that the generated photons quickly leave and associate an \textit{L}-operator given by $L_{e5}=a_{e5}$ and the corresponding loss rate $\gamma_{e5}=2\Gamma_{e5}$
\footnote{In general, one can associate both \textit{loss} and \textit{gain}; balanced in the case of stimulated processes (as in the case of $\mathcal{PT}$-symmetric systems \cite{Bender2007}) or strongly imbalanced in the case of spontaneous transitions.}. 
Furthermore, anticipating that for the photons we will only need the expectation value of the photon operators, we can replace the Lindblad dynamics by that of an effective non-Hermitian Hamiltonian as introduced above. Namely, we modify the second term of the photon Hamiltonian as follows,
\begin{equation}
H_{\mathrm{ph}}=\hbar\omega_{eg}\,a_{eg}^{\dagger}a_{eg}+
\hbar\left(\omega_{e5}-\mathrm{i}\Gamma_{e5}\right)\,
a_{e5}^{\dagger}a_{e5}%
\end{equation}
giving the $\omega_{e5}$ photon mode an intrinsic lifetime.


To highlight the resonant nature of the lambda system, we resort to a series of generalized time-dependent gauge transformations. For added clarity, we proceed in multiple steps. We first concern ourselves with the bosons and perform the formal (rotating-frame) replacements,
\begin{equation*}
a_{eg}\rightarrow 
e^{-\mathrm{i}\omega_{eg}t}a_{eg}
\qquad\text{and}\qquad
a_{e5}\rightarrow 
e^{-\mathrm{i}\left(\omega_{e5}-\mathrm{i}\Gamma_{e5}\right)t}a_{e5}
\end{equation*}
The first one is standard 
\footnote{Using the usual replacement notation for field redefinitions, where $\psi\rightarrow f\left[\psi\right]$ means $\psi=f\left[\psi^{\prime}\right]$ and we subsequently drop the primes.}
and we transform $a_{eg}^{\dagger}$ with the corresponding Hermitian-conjugate relation. But the second one goes \emph{beyond} a pure-phase modulation and, in accordance with Eq.~(\ref{Eq:Ldagger}), we combine it with
\begin{equation}
a_{e5}^{\dagger}\rightarrow e^{\mathrm{i}\left(  \omega_{e5}-\mathrm{i}%
\Gamma_{e5}\right)  t}a_{e5}^{\dagger}
\end{equation}
This choice guarantees that diagonal (chemical) potential terms are unaffected by the transformation.

Due to their explicit time dependence, these transformations need to be done at the level of the Lagrangian,
\begin{align}
L_{\mathrm{ph}} & =
a_{eg}^{\dagger}\left(\mathrm{i}\hbar\partial_{t}\right)a_{eg}+
a_{e5}^{\dagger}\left(  \mathrm{i}\hbar\partial_{t}\right)a_{e5}-
H_{\mathrm{ph}}\notag\\
& \rightarrow a_{eg}^{\dagger}
\left(\mathrm{i}\hbar\partial_{t}\right)a_{eg}+
a_{e5}^{\dagger}\left(\mathrm{i}\hbar\partial_{t}\right)a_{e5}
\end{align}
We remark that in the Lagragian one deals with $H_{\mathrm{ph}}$, rather than its Hermitian conjugate, since the time derivatives act only on annihilation operators.
The net effect of the first set of transformations was to bring to zero, $H_{\mathrm{ph}}\rightarrow0$, the photon term of the Hamiltonian.

In the same vein, we perform similar (standard rotating frame) replacements for the fermion operators,
$c_{n}\rightarrow e^{-\mathrm{i}\varepsilon_{n}t/\hbar}c_{n}$ where $n\!\in\!\left\{g,e,5\right\}$, to readily find $H_{\mathrm{at}}\rightarrow0$. 
The non-trivial nature of the dynamics is now entirely encoded in the interaction part of the Hamiltonian. 
Let us transform and write its terms out explicitly,
\begin{align}
H_{\mathrm{int}}\;\rightarrow\;  
\lambda_{eg}\,e^{-\mathrm{i}\delta_{eg}t}
c_{e}^{\dagger}a_{eg}c_{g}+&
\lambda_{eg}^{\ast}\,e^{\mathrm{i}\delta_{eg}t}
c_{g}^{\dagger}a_{eg}^{\dagger}c_{e}\notag\\
+\,\lambda_{e5}\,e^{-\Gamma_{e5}t}
c_{e}^{\dagger}a_{e5}c_{5}+&
\lambda_{e5}^{\ast}\,e^{\Gamma_{e5}t}
c_{5}^{\dagger}a_{e5}^{\dagger}c_{e}
\end{align}
where we defined $\hbar\delta_{eg}=\hbar\omega_{eg}-
\left(\varepsilon_{e}-\varepsilon_{g}\right)$, with 
$\left\vert\delta_{eg}\right\vert\ll
\left\vert\varepsilon_{e}-\varepsilon_{g}\right\vert$, to model a possible small detuning of the proving laser from the exact transition
frequency.
On the other hand, we will assume that $\hbar\omega_{e5}=\left(\varepsilon_{e}-\varepsilon_{5}\right)$ to a very good precision, since those photons are emitted spontaneously.


At this point, we selected the framework with the ideal conditions to approximate the state of the photons by coherent states:
$\left\vert \alpha\right\rangle=
e^{\alpha a^{\dagger}-\alpha^{\ast}a}\left\vert 0\right\rangle$,
such that $a\left\vert\alpha\right\rangle=
\alpha\left\vert\alpha\right\rangle$ and 
$\left\langle\alpha\vert\alpha\right\rangle=1$,
(giving an overcomplete basis). 
We can now do a mean-field approximation, simply replacing the photon fields by their corresponding time-independent expectation values, 
$a_{m}\rightarrow\left\langle a_{m}\right\rangle=\alpha_{m}$ 
and 
$a_{m}^{\dagger}\rightarrow\left\langle a_{m}^{\dagger}\right\rangle= \alpha_{m}^{\ast}$ for $m\in\left\{eg,e5\right\}$.
The dynamics of the latter was captured \textit{exactly} by the effective non-Hermitian Hamiltonian (and made explicit by the generalized gauge transformations) because they coincide with the \textit{L}-operators of the leaky-cavity model.

The bosons removed from the problem, we are left with a quadratic Hamiltonian for the \textit{internal transitions} in a lambda system described purely by fermions with overlap matrix elements $t_m=\lambda_m\alpha_m$. In the experimental setup, this is further embedded in a junction.


We next resort to a further series of gauge transformations to eliminate the explicit time dependence from the Hamiltonian (cf.~Ref.~\onlinecite{ShahBolech_TransportPuzzle,*ShahBolech_NoneqToulousePoint,LjepojaBolechShah_part1,*LjepojaBolechShah_part2,*LjepojaBolechShah_part3}) moving it towards state $\left\vert 5\right\rangle$. We start by undoing the state-$\left\vert g\right\rangle$ transformation, $c_{g}\rightarrow e^{\mathrm{i}\varepsilon_{g}t/\hbar}c_{g}$.
To continue, we remove the oscillatory time dependence from the $g$-$e$
transition via 
$c_{e}\rightarrow e^{-\mathrm{i}\left(\delta_{eg}-\varepsilon_{g}/\hbar\right)t}c_{e}$. 
Thirdly, we remove the time dependence from the $e$-$5$ transition using
$c_{5}\rightarrow e^{-\mathrm{i}\left(\delta_{eg}-\varepsilon_{g}
/\hbar+\mathrm{i}\Gamma_{e5}\right)t}c_{5}$.
The internal-transitions term is now, as desired, time-independent (and Hermitian),
\begin{equation}
H_{\mathrm{int}}=t_{eg}\,c_{e}^{\dagger}c_{g}+t_{eg}^{\ast}\,c_{g}^{\dagger
}c_{e}+t_{e5}\,c_{e}^{\dagger}c_{5}+t_{e5}^{\ast}\,c_{5}^{\dagger}c_{e}%
\end{equation}
while the atomic Hamiltonian acquires a non-Hermitian third term \cite{Tao2025},
\begin{equation*}
H_{\mathrm{at}}=
\varepsilon_{g}\,c_{g}^{\dagger}c_{g}+
\left(\varepsilon_{g}-\hbar\delta_{eg}\right)c_{e}^{\dagger}c_{e}+
\left(\varepsilon_{g}-\hbar\delta_{eg}-
\mathrm{i}\hbar\Gamma_{e5}\right)c_{5}^{\dagger}c_{5}
\end{equation*}

Next, we highlight that any further connection of state $\left\vert 5\right\rangle$ to other states has now the time dependence given by the net transformation $c_{5}\rightarrow e^{-\mathrm{i}\left[\delta_{eg}+\left(\varepsilon_{5}-\varepsilon_{g}\right)  /\hbar+\mathrm{i}\Gamma_{e5}\right]t}c_{5}$.


In the experiment, atoms in the state $\left\vert5\right\rangle$ are not trapped and quickly leave the system without further interactions. 
We shall thus model that state as connected to an \textit{empty} fermionic bath (with an infinitely low chemical potential), and we need to deal with the remaining explicit time dependence still present in that connection.

Let us introduce the state $\left\vert5b\right\rangle $ as the first one encountered after the \emph{leap} of state $\left\vert 5\right\rangle$ into the \emph{bath}, and say it is part of a lattice with all site energies equal to $\varepsilon_{5}$. 
Consider, for simplicity, some first-neighbor hopping lattice and the hopping between the two sites in question is
\begin{align}
t_{5}\,c_{5b}^{\dagger}c_{5}+t_{5}^{\ast}\,c_{5}^{\dagger}c_{5b}
\rightarrow
t_{5}\,e^{-\mathrm{i}\left[\delta_{eg}+
\left(\varepsilon_{5}-\varepsilon_{g}\right)/\hbar+
\mathrm{i}\Gamma_{e5}\right]t}c_{5b}^{\dagger}c_{5}\notag\\+
t_{5}^{\ast}\,e^{\mathrm{i}
\left[\delta_{eg}+\left(\varepsilon_{5}-\varepsilon_{g}\right)/\hbar+
\mathrm{i}\Gamma_{e5}\right]t}c_{5}^{\dagger}c_{5b}    
\end{align}
One can remove the time dependence that was introduced by the transformation of $c_{5}$ by identically transforming $c_{5b}$; and so on for the entire bath lattice.


Let us model the state-$\left\vert 5b\right\rangle $ bath as a continuum linearized band. 
Then we can identify $c_{5b}\equiv\psi_{5b}\left(x=0,t\right)$ and the corresponding Lagrangian for the entire bath is
$\psi_{5b}^{\dagger}\left(x,t\right)
\left(\mathrm{i}\hbar\partial_{t}+
\mathrm{i}\hbar v_{\text{\textsc{f}}}\partial_{x}\right)
\psi_{5b}\left(x,t\right)$; where we chose to use right-moving fermions. 
We can now generalize the gauge transformation and make it space-time dependent \cite{ShahBolech_TransportPuzzle,*ShahBolech_NoneqToulousePoint},
\begin{equation}
\psi_{5b}\left(x,t\right)\rightarrow 
e^{-\mathrm{i}\left[\delta_{eg}+
\left(\varepsilon_{5}-\varepsilon_{g}\right)/\hbar+
\mathrm{i}\Gamma_{e5}\right]\left(t-x/v_{\text{\textsc{f}}}\right)}
\psi_{5b}\left(x,t\right)
\end{equation}
Notice that at $x\!=\!0$ the transformation is the same as before, while into the lead it changes continuously. 
This has the net effect of bringing back the background energy to its original value, $\varepsilon_{5}$, but also introduces an effective complex shift in the chemical potential $\delta\mu_{\mathrm{eff}}=\hbar\delta_{eg}+
\left(\varepsilon_{5}-\varepsilon_{g}\right)+\mathrm{i}\Gamma_{e5}$, so that $\mu_{\mathrm{eff}}=\mu_{5b}+\delta\mu_{\mathrm{eff}}$.

If we now consider bath $\left\vert 5b\right\rangle$ as decoupled from any other states and integrate its different Green functions over momentum to derive a \emph{local action} for its site at $x=0$ \cite{bolech_tunnelling_short,*bolech_tunnelling_long,BolechDemler_MajoranaNoise}, we find in the (reordered) \emph{Keldysh basis} (Kb) \cite{keldysh_method,Visuri2023,ThompsonKamenev_LindbladianDynamics}:
\begin{equation*}
\left.\mathrm{i}G_{5b}^{-1}\left(\omega\right)\right\vert_{\mathrm{Kb}}=
2v_{\text{\textsc{f}}}
\begin{pmatrix}
0 & 1\\
-1 & -2s_{\text{\textsc{f}}}\left(\omega\right)
\end{pmatrix}
\rightarrow2v_{\text{\textsc{f}}}
\begin{pmatrix}
0 & 1\\
-1 & -2
\end{pmatrix}
\end{equation*}
where $v_{\text{\textsc{f}}}$ is the bath's Fermi velocity and
\begin{equation}
s_{\text{\textsc{f}}}\left(\omega\right)=
\tanh\left(\frac{\hbar\omega-\mu_{\mathrm{eff}}}{2k_{\text{\textsc{b}}}T}\right)\rightarrow
\lim_{\mu_{5}\rightarrow\left(-\infty\right)}
s_{\text{\textsc{f}}}\left(\omega\right)=1
\end{equation}
Notice that in this {\textquotedblleft standard Marcovian limit\textquotedblright} the effect of the chemical potential shift is lost.
We remark that this happens even if the shift has an imaginary part and that is the reason we do not treat state-$\left\vert5\right\rangle$ itself as part of the bath, for information about the outgoing photons would have been lost.


\begin{figure}[t]
\begin{center}
\includegraphics[width=\columnwidth]{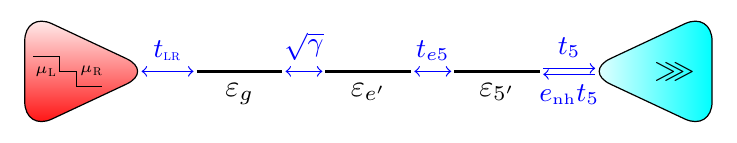}
\end{center}
\caption{\label{Fig:effective-model}
Schematic depiction of the purely fermionic effective model of a lambda system carrying a \textit{loss current} between a two-lead junction (left side) and a Markovian reservoir (right side). The driving-laser intensity is $\gamma$, the effective site energies are real-valued and shifted by the detuning from the $g$-$e$ transition, $\varepsilon_{e'}=\varepsilon_{5'}=\varepsilon_g-\hbar\delta_{eg}$. The connection to the Markovian bath is \textit{asymmetric}, with the return amplitude enhanced by the factor $e_\textrm{nh}$.}
\end{figure}

After absorbing state $\left\vert5b\right\rangle$ into a continuum of outgoing states (modeled by a local action, $\mathrm{i}G_{5b}^{-1}$), we can keep only its \emph{effective hybridization} with state $\left\vert5\right\rangle$.
Integrating out the bath completely, one arrives at a non-Hamiltonian description in terms of a Keldysh action equivalent to treating the bath as an \textit{L}-operator, $L_{5}=c_{5}$, that models atom loss directly from state $\left\vert5\right\rangle$ with loss rate 
$\gamma_{5}=2\Gamma_{5}=2\left\vert t_{5}\right\vert^{2}/\hbar$.

The net effect of the bath coupling on the Keldysh action for state $\left\vert5\right\rangle$ is thus to further shift its site energy.
If we introduce the dimensionless \textit{enhancement} parameter $e_{\mathrm{nh}}=\allowbreak1+\Gamma_{e5}/\Gamma_{5}\geq\allowbreak1$, the resulting expression is
\begin{equation}
\varepsilon_{5}\rightarrow
\left(\varepsilon_{g}-\hbar\delta_{eg}\right)-
\mathrm{i}\hbar e_{\mathrm{nh}}\Gamma_{5}
\end{equation}

Based on prior calculations using a time-loop Keldysh formalism extension to (asymmetric) non-Hermitian systems \cite{KakashviliBolech_TimeLoop}, we can match expressions to see that the case of $e_\textrm{nh}>1$ corresponds to a \emph{nonreciprocal} connection between $5\rightleftarrows5b$ (see Fig.~\ref{Fig:effective-model}),
\begin{equation}
H_{\mathrm{55b}}=
t_{5}\,c_{5b}^{\dagger}c_{5}+
e_{\textrm{nh}}t_{5}^{\ast}\,c_{5}^{\dagger}c_{5b}
\end{equation}
Counterintuitively, it is the \emph{reverse} transition that is enhanced; but this is consistent with a \emph{suppression} of the reverse $e\to5$ transition expected on physical grounds due to photon loss.
Notice that in the effective theory corresponding to Fig.~\ref{Fig:effective-model}, the ``non-Hermiticity'' appears in a two-fold way: (i) in the integration of the Gibbsian and pure-Markovian reservoirs, ---for the transport and the loss to state $5b$, respectively; (ii) in the nonreciprocal terms appearing in the connection of the system to the latter reservoir (stemming from the elimination of the photons). 
Remark that these nonreciprocal terms do not need post-selection of data, nor the absence of quantum jumps on the trajectories. 
One can thus \emph{a priori} expect observable effects due to the combined presence of both of these aspects in the effective Keldysh action. 

\begin{figure}[ht]
\begin{center}
\includegraphics[width=\columnwidth]{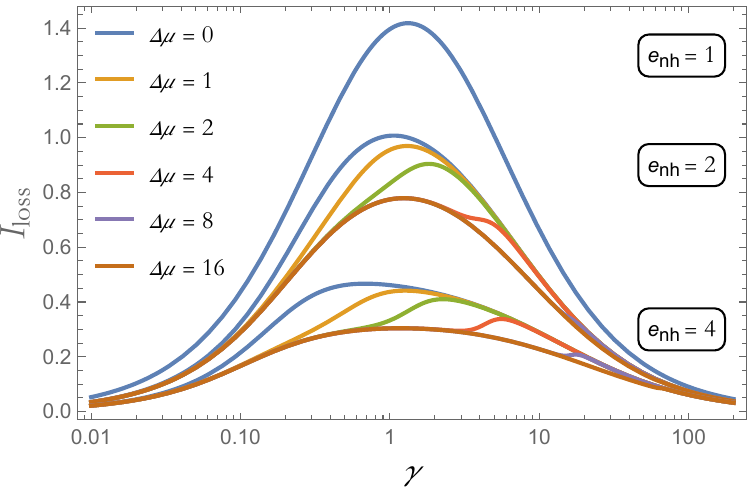}
\end{center}
\caption{\label{Fig:LossCurrent}
Loss current as a function of driving-laser intensity ($\gamma$). Three sets of curves are shown for different values of the enhancement parameter. When $e_\textrm{nh}\!=\!1$, there is no dependence on the chemical potential drop across the junction, ($\Delta\mu\!=\!\mu_\textsc{l}\!-\!\mu_\textsc{r}\!=\!2\mu_\textsc{l}$ with $\varepsilon_g\!=\!\delta_{eg}\!=\!0$), and all curves are superimposed. For $e_\textrm{nh}\!>\!1$, the shape of the qZE is affected and the curves cross over in a $\Delta\mu$-dependent way between the extreme non-equilibrium case ($\Delta\mu\to\infty$ \cite{Brandes2005,ContrerasPulido2013}) at low $\gamma$ and the equilibrium-junction curve ($\Delta\mu=0$) at large $\gamma$. Here we used units such that $2v_\textsc{f}\!=\!1$, the temperature is $k_\textsc{b}T\!=\!0.1$ and we choose $t_\textsc{l,r}\!=\!1/2$ and $t_{e5,5}\!=\!1$.}
\end{figure}

In order to test for this, let us compute the loss-current, $I_\textrm{loss}$, corresponding to the atoms excited by the laser light into 
state $e$ (topping the lambda system) that ultimately end up exiting the trap, as shown in Fig.~\ref{Fig:LossCurrent}. 


The current first increases with $\gamma$, but for sufficiently large values it saturates and starts to decrease; a feature known as the continuous ``quantum Zeno effect'' (qZE) \cite{Daley2014_QuantumTrajectories,sieberer2023_UniversalityDrivenOpenQuantum}. Remarkably, driving the junction out of equilibrium can be used to reveal the nonreciprocitiy of the effective model (for $e_\text{nh}\!>\!1$), which is a signature of the dual loss mechanism (atoms and photons) responsible for the \emph{enhanced irreversibility} of the process
\footnote{Notice that the extreme non-equilibrium junction at any finite temperature is always equivalent to an effective single Gibbsian reservoir in the infinite-temperature limit. Given this, it could be interesting to explore the possible appearance of \emph{exceptional points} that has been found in the study of quantum thermal machines \cite{KhandelwalHaack_QuantumThermalMachine}.}.


In summary, we presented a careful description of the dissipative nature of spontaneous emission that is well adapted to the description of continuous processes and can be easily combined with additional dissipation mechanisms. For the case we studied, this goes beyond the joint treatment of the dual loss process via a single \textit{L}-operator, $L_{e5+5}\!=\!a_{e5}c_{5}$ (which would miss the possible interplay of two different time scales; cf.~with considering only atom loss \cite{Corman_DissipativeAtomicPointContact,Uchino2022,Visuri2023_DCtransport}). We have shown there are observable consequences if the dissipation is engineered over steady states of systems driven out of equilibrium. Moreover, it is possible to control the value of $e_\textrm{nh}$ (in particular, lowering it and switching from $>$$1$ to $1$) by adding a secondary laser that (weakly) drives the $e$-$5$ transition; cf.~Ref.~\onlinecite{Talebi_DarkStateTransport}. The ubiquitous use of three-state transition systems (\emph{lambda}, \emph{vee}, or \emph{cascade}) in numerous optics experiments \cite{MinevDevoret2019,NaghilooJoglekarMurch2019,*ChenJoglekarMurch2022,*AbbasiJoglekarMurch2022,*ChenJoglekarMurch2022,*ErdamarJoglekarMurch2024,Tao2025} suggests that our ideas can also be applicable to many other interesting situations. In the solid-state context, similar considerations would apply also to phonon relaxation processes, to give just one example.

\begin{acknowledgments}
We thank T.~Esslinger, M.-Z.~Huang, M.~Talebi, S.~Wili and Y.~Yudkin for interesting discussions on their experiments and A.~Daley, S.~Diehl, J.M.~Raimond and L.~Sanchez-Palencia for interesting exchanges on the theory of open systems. This work was supported in part by the Swiss National Science Foundation under Division II (Grant No. 200020-219400).
\end{acknowledgments}

%

\end{document}